# Estimation of genome size using *k*-mer frequencies from corrected long reads

Hengchao Wang, Bo Liu, Yan Zhang, Fan Jiang, Yuwei Ren, Lijuan Yin, Hangwei Liu, Sen Wang and Wei Fan*

Agricultural Genomics Institute at Shenzhen, Chinese Academy of Agricultural Sciences, Shenzhen, Guangdong, 518120, China. *e-mail: fanwei@caas.cn

**Abstract:** The third-generation long reads sequencing technologies, such as PacBio and Nanopore, have great advantages over second-generation Illumina sequencing in *de novo* assembly studies. However, due to the inherent low base accuracy, third-generation sequencing data cannot be used for *k*-mer counting and estimating genomic profile based on *k*-mer frequencies. Thus, in current genome projects, second-generation data is also necessary for accurately determining genome size and other genomic characteristics. We show that corrected third-generation data can be used to count *k*-mer frequencies and estimate genome size reliably, in replacement of using second-generation data. Therefore, future genome projects can depend on only one sequencing technology to finish both assembly and *k*-mer analysis, which will largely decrease sequencing cost in both time and money. Moreover, we present a fast light-weight tool kmerfreq and use it to perform all the *k*-mer counting tasks in this work. We have demonstrated that corrected third-generation sequencing data can be used to estimate genome size and developed a new open-source C/C++ *k*-mer counting tool, kmerfreq, which is freely available at https://github.com/fanagislab/kmerfreq.

In recent years, third-generation sequencing (TGS) technologies, such as Pacific Biosciences (PacBio), have become dominant in *de novo* assembly of large genomes. TGS is typically known as real-time single-molecule sequencing, which uses native DNA fragments to sequence instead of template amplification, avoiding copying errors, sequence-dependent biases and information losses(1). The development of TGS assembly algorithms is also blooming, and several TGS assembly tools such as HGAP(2), Canu(3), Falcon(4), Miniasm(5), MECAT(6), wtdbg2(7) and Flye(8), have become mature and been widely adopted. Utilizing advantages of even genomic coverage and ultra-long read length, continuity of TGS assembly has surpassed any sequencing technology developed before.

Although assembly quality has been largely improved with TGS, until now, there is no assembler that can generate complete genome for large plant or animal genomes with complex structure, and most genome projects still need to perform an assembly-independent estimation of genomic characteristics based on *k*-mer frequencies. With *k*-mer frequencies counted from raw reads, probability models can be built to estimate genome size, repeat content, and heterozygous rate(9, 10). In addition, it has been reported that accuracy of this method is higher than that of traditional golden standard with DNA flow cytometry(11), making it a necessary and valuable analysis in *de novo* genome studies.

However, due to inherent low base accuracy(2, 12, 13), TGS data cannot be used for *k*-mer counting and estimating genomic profile based on *k*-mer frequencies. In fact, *k*-mer related analysis has always been dominated by second-generation sequencing (SGS) data, especially Illumina. Thus, in current genome projects, SGS data is also necessary for accurately estimating genome size and other genomic characteristics, increasing cost of both time and money. Even though raw TGS reads have much lower base accuracy than Illumina raw reads, the accuracy of TGS reads could be significantly improved by error correction, utilizing the multiplicity and consensus information in TGS data(14).

Here, we propose a *k*-mer frequency based genome size estimation method with corrected TGS reads, in replacement of using SGS reads traditionally. To confirm the feasibility, we tested this method on both simulated and real data. Moreover, we present a light-weight *k*-mer counting program, kmerfreq, to facilitate *k*-mer counting in this work.

## Methods and Results

### Overview and working principle of kmerfreq

*k*-mer counting that aims to determine the frequency of *k*-length substrings (*k*-mers) in the sequencing data, is a basic tool for *k*-mer frequency based estimation and also a frequent job in many bioinformatics applications. There are several available tools, which can be classified into 4 major types by underlying algorithms, including hash table, sorting, burst tries and enhanced suffix array(15). Although the efficiency of recently published tools such as KMC3 have been largely improved, counting *k*-mers from large amount of reads data is still not a trivial task. For genomic characteristics estimation, *de novo* genome studies often use smaller *k*-mer size, mostly 17. One reason is that the total *k*-mer space ($4^{17}$=16 G) is enough larger than the genome size of most common genomes and thus has the ability to store all the *k*-mers derived from the genomes; Another reason is that using a larger *k*-mer size will result in more erroneous *k*-mers caused by sequencing errors and then decrease efficiency of this method. In other words, the higher error rate in the sequencing data, the smaller *k*-mer size should be used. For this purpose, we developed a fast light-weight tool, kmerfreq, to perform *k*-mer counting specialized for smaller *k*-mer sizes (< 19), and use it to count all the *k*-mer frequencies shown in this work.

kmerfreq operates with a fixed memory size in parallel computational mode. It adopts a dynamic array to store the frequency value of all potential *k*-mers with size k, using two-bytes to store each frequency value, and taking the *k*-mer bit-value converted from the *k*-mer sequence as index of the frequency array, so the total memory usage is $2\times4^k$ bytes. It uses the main thread to load data from disk into memory, and multiple children threads to count *k*-mers frequency with lock-free CAS (compare and swap) operations simultaneously. Moreover, the *k*-mer chopping and bit-value converting method are also optimized by exploiting the property that two successive *k*-mers share a (k -1) bases to enhance speed efficiency.

We compared the performance of kmerfreq with other published tools at *k*-mer size 17, using 30 X (90 G) human Illumina data simulated by pIRS(16). kmerfreq uses moderate 32 G memory, and its speed is much higher than most published tools and commensurate with that of KMC3(17), which has been the fastest *k*-mer counting tool available (Table 1 and Supplementary Table 1). It is worth to note that kmerfreq does not utilize disk to process temporary results and do not output any middle result into disk. As a result, it has higher disk efficiency than KMC3 and other tools, and is more convenient for users.

**Genome size estimation for synthetic datasets**

The distribution of *k*-mer frequencies is highly influenced by the level of sequencing errors. As we known, there is a huge gap between accuracy level of TGS and SGS data, and it is not sure what level of base accuracy could be suitable for *k*-mer frequency based estimation. Theoretically, if signal-to-noise ratio is high enough, i.e. if peaks that formed by random sampling of genomic *k*-mers could be clearly observed from *k*-mer frequency curve, then *k*-mer frequency based estimations would work. To evaluate, we simulated a set of human PacBio data with gradient accuracy by PBSIM(18) and counted *k*-mer frequencies (k = 17) independently. From the distribution curves (Fig. 1a), we can infer, when base accuracy level is higher than 96%, sequencing data could be potentially used for *k*-mer frequency based estimations. Further analysis confirmed this hypothesis and showed that the higher base accuracy, the more accurate genome size estimation (Supplementary Table 2). Furthermore, we also confirmed that PacBio circular consensus sequencing (CCS) reads with over 99% average base accuracy can be used to estimate genome size with high accuracy (Supplementary Fig. 1, Supplementary Table 2 and 6). On the contrary, current TGS data only has 85% base accuracy, and error correction that aims to increase accuracy level must be performed, in order to use TGS data for *k*-mer frequency based estimations. Luckily, there are already several available tools for TGS error correction, such as the built-in tool in Canu package(3).

Firstly, we applied this method to simulated data of 9 model species, including 1 bacterium (*Escherichia coli*), 1 fungus (*Saccharomyces cerevisiae*), 5 animals (*Caenorhabditis elegans*, *Drosophila melanogaster*, *Danio rerio*, *Gallus Gallus*, *Homo sapiens*), and 2 plants (*Arabidopsis thaliana*, *Oryza sativa*) (Supplementary Table 3). Considering no heterozygosity, Illumina data were simulated by pIRS, PacBio data were simulated by PBSIM (Supplementary Table 4), and corrected PacBio data were generated by a built-in tool in Canu. On distribution curves of *k*-mer frequency, there is no obvious peak for raw PacBio data, in contrast, clear peaks could be observed for corrected PacBio data, similar to that of Illumina data (Fig. 1b and Supplementary Fig. 2). Genome size estimations with GCE method(9) and a naive method that directly uses total *k*-mer individuals divided by the major peak value, both produced highly accurate genome sizes, almost all with error rate less than 3%, and the difference between using corrected PacBio and Illumina data is also less than 3% (Supplementary Table 5), suggesting that this method is highly accurate for simulated data.

### Genome size estimation for genuine datasets

Next, we extended this method to real sequencing data of 11 species, including 6 animals (*Caenorhabditis nigoni*, *Taenia multiceps*, *Pyrocoelia pectoralis*, *Sillago sinica*, *Chaenocephalus aceratus*, *Aedes aegypti*) and 5 plants (*Durio zibethinus*, *Oropetium thomaeum*, *Mikania micrantha*, *Panicum miliaceum*, *Cinnamomum kanehirae*) (Supplementary Table 6 and 7). Besides sequencing errors, real sequencing data also has heterozygosity and coverage bias problems, which complicated *k*-mer frequency curves. As expected, all curves of PacBio data do not show any peak, while corrected PacBio curves show similar peaks to that of Illumina data (Fig. 1c and Supplementary Fig. 3). However, there is a bigger difference on the curve shapes for real data than simulated data. Genome size estimations with naive and GCE method also showed larger difference between using corrected PacBio and Illumina data, but for most genomes corrected PacBio data can produce reliable genome size estimations (Supplementary Table 8), except for some genomes with extremely complex structure, such as *Durio zibethinus* and *Mikania micrantha*, due to high heterozygous rate, suggesting that this method is also feasible for real sequencing data of most species.

## Discussion

The cost and throughput of TGS technologies have been dramatically improved(19), approaching to that of SGS technologies. Thus, large-scale replacement of SGS with TGS occurs in most bioinformatics applications, especially in *de novo* genome studies. Although assembly continuity has been greatly enhanced with TGS data, *k*-mer frequency based estimations can still provide a beneficial supplementary information for genome profile. Here, we only tested with PacBio technology, but there should be no problem to extend the method to Oxford Nanopore Technologies (ONT) or other TGS technologies. The key and speed-limited step is error correction of TGS data, however, this situation may be changed by the endeavor to increase base accuracy of TGS, such as CCS achieved by Sequel System 6.0, and ONT R10 with a longer barrel and dual reader head. With increased base accuracy of raw data, it will become easier for error correction of TGS data, making *k*-mer frequency based estimation more practicable and accurate in the near future.

## Conclusion

We proposed and demonstrated that corrected TGS data can be applied for *k*-mer counting and genomic profile estimation, in replacement of using SGS data traditionally. Therefore, future genome projects can depend on only one TGS technology to finish both major assembly and supplementary *k*-mer frequency analysis, which will largely decrease sequencing cost in both time and labour. Moreover, we present a fast light-weight *k*-mer counting tool, kmerfreq, which is comparable in speed to the fastest *k*-mer counting tool available and also have several additional advantages to facilitate genome estimation.

## Acknowledgements

The work was supported by Shenzhen science and technology program (JCYJ20170303154245825); The Agricultural Science and Technology Innovation Program Cooperation and Innovation Mission (CAAS-XTCX2016).

## Author contributions

The study was designed by W. F. H. W. performed benchmark of *k*-mer counting tools and wrote draft manuscript. B. L., Y. Z. and F. J. downloaded real sequencing data in published papers. Y. R., L. Y., H. L. and S. W. downloaded model genomes for simulation. All co-authors substantively revised the manuscript and approved the submitted version.

## Competing interests

The authors declare no competing interests.

**Table 1.** *k*-mer counting performance of various tools.

| Softwares | Wall time (s) | Max memory (G) | Disk space (G) |
|---|---|---|---|
| kmerfreq | 1,887 | 32 | 0.0032 |
| KMC3 | 1,948 | 30 | 12 |
| Jellyfish | 7,024 | 30 | 24 |
| DSK | 3,283 | 25 | 43 |
| KCMBT | 3,016 | 101 | 22 |
| BFCounter | 65,061 | 22 | 51 |
| GenomeTester4 | 4,540 | 192 | 32 |
| tallymer | 116,361 | 199 | 842 |
| KAnalyze | 348,912 | 10 | 55 |

All tools are run in triplicate and the average of each feature are shown, except KAnalyze, which run one time, because it takes too much time to run and we suppose its statistics variation should be small. All the experiments, if not specifically mentioned, are conducted by 4 threads on a server with 2.00GHz Intel Xeon CPU E7-4830 v4 on CentOS release 6.6 system. We compared the performance of kmerfreq with other published tools at *k*-mer size 17, using 30 X (90 G) human Illumina data simulated by pIRS.

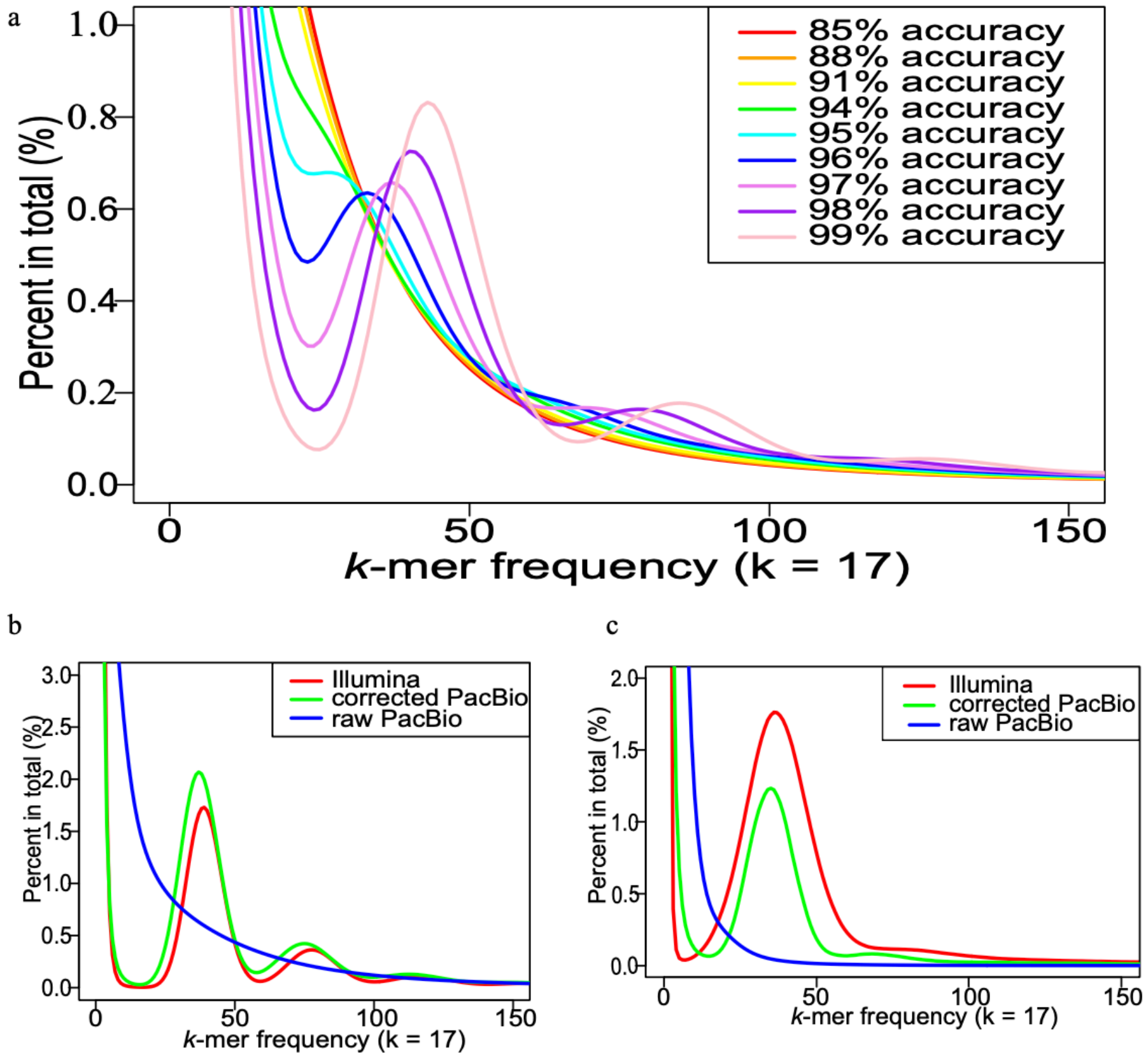

**Figure 1. The curves of *k*-mer distribution.** (a) The distribution of *k*-mer frequencies from 50 X simulated PacBio datasets of *Homo Sapiens* with different accuracy. (b) The distribution of *k*-mer frequencies from 45 X simulated Illumina, corrected PacBio and 100 X raw PacBio datasets of *Homo Sapiens*. (c) The distribution of *k*-mer frequencies from 40 XIllumina, corrected PacBio and 125 X raw PacBio datasets of *Taenia multiceps*.

# Supplementary materials

## 1. Supplementary methods

### 1.1 *k*-mer counting method

kmerfreq counts *k*-mer (with size *k*) frequency from the input sequence data, typically sequencing reads data, and reference genome data is also applicable. The forward and reverse strand of a *k*-mer are taken as the same *k*-mer, and only the *k*-mer strand with smaller bit-value is used to represent the *k*-mer. It adopts a 16-bit integer with max value 65535 to store the frequency value of a unique *k*-mer, and any *k*-mer with frequency larger than 65535 will be recorded as 65535. The program stores all *k*-mer frequency values in a 4^*k* size array of 16-bit integer (2 bytes), using the *k*-mer bit-value as index, so the total memory usage is 2*4^*k* bytes. For *k*-mer size 15, 16, 17, 18, 19, it will consume constant 2G, 8G, 32G, 128G, 512G memory, respectively. kmerfreq works in a highly simple and parallel style, to achieve as fast speed as possible.

One of noteworthy method is *k*-mer bit calculation, which exploits the property that two successive *k*-mers share a (*k* -1) bases, to transfer read sequence into *k*-mer bit value. For each read, the first *k*-mer sequence is chopped and transferred into bit value. For next *k*-mer in this read, we do not chop sequence and only need to transfer the newly added base into bit value and combine with the last bit value with bit operations.

In summary, kmerfreq has some important features:

(1) Fixed memory: stores all *k*-mer frequency values in a 4^*k* size array of 16-bit integer (2 bytes), using the *k*-mer bit-value as index. adopts a 16-bit integer with max value 65535 to store the frequency value of a unique *k*-mer.

(2) Highly parallel: main thread load data, while the children threads simultaneously count *k*-mers without waiting, using lock free technology as Jellyfish.

(3) *k*-mer operation: advanced method of chopping reads into *k*-mers, and converting *k*-mer sequence to *k*-mer bit values.

(4) Diverse output: Distribution of *k*-mer frequency (default), *k*-mer sequence and frequency values with a cutoff (optional), compressed data structure [one-bit for a unique *k*-mer, 0 for low-frequency, 1 for high-frequency] (optional)

## 1.2 Information for the *k*-mer counting tools

### 1.2.1 Download addresses:

| | | | |
|---|---|---|---|
| kmerfreq | 4.0 | ceffd13 | https://github.com/fanagislab/kmerfreq |
| KMC3 | 3.1.0 | 4287983 | https://github.com/refresh-bio/KMC |
| Jellyfish | 2.2.7 | 2cff4e4 | https://github.com/gmarcais/Jellyfish |
| DSK | 2.1.0 | | https://github.com/GATB/dsk |
| KAnalyze | 2.0.0 | | https://sourceforge.net/projects/kanalyze/files/v2.0.0/ |
| KCMBT | 1.0.0 | b4d684a | https://github.com/abdullah009/kcmbt_mt |
| GenomeTester4 | 4.0 | a9c14a6 | https://github.com/bioinfo-ut/GenomeTester4 |
| BFCounter | 1.0 | f7f96a1 | https://github.com/pmelsted/BFCounter |
| genometools-1.5.10 (tallymer) | 1.5.10 | | http://genometools.org/pub/ |
| runit.pl | 5e577d7 | | https://github.com/ruanjue/wtdbg2/blob/master/scripts/runit.pl |
| Canu | 1.8 | 22a2453 | https://github.com/marbl/canu |

### 1.2.2 Command lines used for comparison:

- kmerfreq

./kmerfreq -k 17 -t 4 -p human_k17_t4 reads.lib

Note that reads.lib includes pair end simulated fastq files.

- kmc3(1)

./kmc -v -k17 -m32 -fq -ci1 -t4 -cs65535 @reads.lib human_k17_t4 ./bins

./kmc_tools transform human_k17_t4 histogram human_k17_t4_histo.txt

- jellyfish(2)

./jellyfish count -m 17 -s 10G -t 4 -o human_k17_t4.jf -C ./*fq

./jellyfish histo -t 4 human_k17_t4.jf > human_k17_t4.jf.histo

- dsk(3)

./dsk -file reads.lib -kmer-size 17 -verbose 2 -nb-cores 4 -out human_k17_t4 -abundance-min-threshold 1 -max-memory 32000 -max-disk 200000 -abundance-min 1

- kcmbt(4)

./kcmbt -k 17 -t 4 -i @reads.lib

./kcmbt_dump 4

Note that reads.lib includes pair end simulated fastq files.

- BFCounter(5)

./BFCounter count -k 17 -t 4 -o human_k17_t4 ./Homo_simulated_illumina_30X_150_350_1.fq ./Homo_simulated_illumina_30X_150_350_2.fq -n 4000000000

./BFCounter dump -k 17 -i human_k17_t4 -o human_k17_t4.dump

- GenomeTester4(6)

./glistmaker ./Homo_simulated_illumina_30X_150_350_1.fq ./Homo_simulated_illumina_30X_150_350_2.fq -w 17 --num_threads 4 -o human_k17_t4

- tallymer(7)

./gt suffixerator -dna -pl -tis -suf -lcp -v -parts 4 -indexname human_30x -db ./Homo_simulated_illumina_30X_150_350_1.fq ./Homo_simulated_illumina_30X_150_350_2.fq

./gt tallymer mkindex -scan -mersize 17 -esa human_30x

- Kanalysis(8)

java -Xmx32G -jar kanalysis-2.0.0/kanalyze.jar count -f fastq -k 17 -o human_k17_t4 -rcanonical -t 4 ./Homo*.fq

## 2. Supplementary tables

### Table 1. Benchmark of k-mer counting tools

| Software | Wall time (s) | User time (s) | System time (s) | CPU time (s) | Maxrss (kb) | Maxvsz (kb) | Disk |
|---|---|---|---|---|---|---|---|
| kmerfreq | 1,887 | 7,025 | 101 | 7,077 | 33,565,732 | 33,883,036 | 3.2M |
| KMC3 | 1,948 | 6,588 | 305 | 6,893 | 31,389,429 | 32,040,304 | 12G |
| Jellyfish | 7,024 | 27,109 | 157 | 26,999 | 31,961,517 | 32,277,624 | 24G |
| DSK | 3,283 | 12,118 | 441 | 12,559 | 25,711,087 | 39,644,856 | 43G |
| KCMBT | 3,016 | 5,832 | 737 | 6,569 | 105,861,517 | 179,434,071 | 22G |
| BFCounter | 65,061 | 123,303 | 207 | 123,510 | 23,185,739 | 23,412,661 | 51G |
| GenomeTester4 | 4,540 | 13,306 | 92 | 13,398 | 200,881,667 | 247,748,893 | 32G |
| tallymer | 116,361 | 113,817 | 2,307 | 116,124 | 208,334,960 | 208,402,364 | 842G |
| KAnalyze | 348,912 | 751,653 | 224,463 | 976,116 | 10,580,400 | 49,545,668 | 55G |

All tools are run in triplicate and the average of each feature are shown, except KAnalyze, which run one time, because it takes too much time to run and we suppose its statistics variation should be small. All the experiments, if not specifically mentioned, are conducted by 4 threads on a server with 2.00GHz Intel Xeon CPU E7-4830 v4 on CentOS release 6.6 system. Maxrss (non-swapped physical memory used). Maxvsz (max virtual size). We compared the performance of kmerfreq with other published tools at *k*-mer size 17, using 30 X (90 G) human Illumina data simulated by pIRS(9).

### Table 2. Genome size estimated by naive method and GCE method

| | genome size | CCS dataset | 96% accuracy | 97% accuracy | 98% accuracy | 99% accuracy |
|---|---|---|---|---|---|---|
| Estimated genome size by naive method | 3,095,950,843 | 3,111,052,276 | 3,925,043,827 | 3,584,550,659 | 3,415,046,927 | 3,269,115,270 |
| Estimated genome size by GCE | | 2,975,070,000 | 4,061,810,000 | 3,814,300,000 | 2,985,250,000 | 3,172,110,000 |

Naive method directly uses total number of *k*-mers divided by the major peak value in the *k*-mer frequency curve.

**Table 3. Model genomes used in this paper**

| Sample | Species | Reference genome |
|---|---|---|
| *H. sap* | *Homo sapiens* | GRCh38.p12 |
| *A. tha* | *Arabidopsis thaliana* | TAIR10.1 |
| *C. ele* | Caenorhabditis elegans | WBcel235 |
| *D. rer* | *Danio rerio* | GRCz11 |
| *D. mel* | *Drosophila melanogaster* | Release 6 plus ISO1 MT |
| *E. col* | *Escherichia coli* | ASM584v2 |
| *G. gal* | *Gallus Gallus* | GRCg6a |
| *O. sat* | *Oryza sativa* | Build 4.0 |
| *S. cer* | *Saccharomyces cerevisiae* | R64 |

The genomic sequences are used to simulate sequencing datasets.

**Table 4. Simulated datasets of model genomes**

| Sample | No. of PacBio reads | No. of PacBio bases | PacBio mean read length | No. of Illumina reads | No. of Illumina bases | Illumina read length |
|---|---|---|---|---|---|---|
| *H. sap* | 31,357,620 | 309,595,203,798 | 8,706 | 928,784,684 | 139,317,702,600 | 150 |
| *A. tha* | 1,209,819 | 11,948,242,700 | 9,911 | 39,827,472 | 5,974,120,800 | 150 |
| *C. ele* | 1,014,756 | 10,028,640,414 | 9,776 | 33,428,798 | 5,014,319,700 | 150 |
| *D. rer* | 16,975,413 | 167,451,426,591 | 9,155 | 558,168,340 | 83,725,251,000 | 150 |
| *D. mel* | 1,595,281 | 14,257,760,263 | 2,217 | 47,522,816 | 7,128,422,400 | 150 |
| *E. col* | 47,086 | 464,165,200 | 9,858 | 1,547,216 | 232,082,400 | 150 |
| *G. Gal* | 10,704,699 | 105,558,209,857 | 6,846 | 351,859,356 | 52,778,903,400 | 150 |
| *O. sat* | 3,774,882 | 37,271,699,547 | 9,372 | 124,238,980 | 18,635,847,000 | 150 |
| *S. cer* | 123,153 | 1,215,713,671 | 9,849 | 4,052,352 | 607,852,800 | 150 |

**Table 5. Genome estimation from simulated datasets of model genomes**

| Sample | genome size | naive method | | GCE method | |
|---|---|---|---|---|---|
| | | Corrected PacBio (% of difference) | Illumina (% of difference) | Corrected PacBio (% of difference) | Illumina (% of difference) |
| *H. sap* | 3,095,950,843 | 3,167,139,320 (2.3%) | 3,114,124,378 (0.5%) | 3,150,720,000 (1.7%) | 3,108,360,000 (0.4%) |
| *A. tha* | 119,482,896 | 119,629,646 (0.1%) | 120,610,585 (0.9%) | 116,810,000 (2.2%) | 119,877,000 (0.3%) |
| *C. ele* | 100,286,401 | 100,535,739 (0.2%) | 101,482,318 (1.1%) | 101,790,000 (1.4%) | 100,877,000 (0.6%) |
| *D. rer* | 1,674,509,851 | 1,717,630,138 (2.6%) | 1,686,372,595 (0.7%) | 1,693,140,000 (1.1%) | 1,687,120,000 (0.8%) |
| *D. mel* | 142,573,024 | 145,725,933 (2.2%) | 143,921,711 (0.9%) | 143,612,000 (0.7%) | 143,157,000 (0.4%) |
| *E. col* | 4,641,652 | 4,785,425 (3.1%) | 4,669,979 (0.6%) | 4,492,770 (3.2%) | 4,653,550 (0.3%) |
| *G. gal* | 1,055,580,959 | 1,081,325,023 (2.4%) | 1,080,021,368 (2.3%) | 1,062,170,000 (0.6%) | 1,059,600,000 (0.4%) |
| *O. sat* | 372,716,981 | 379,811,557 (1.9%) | 378,313,289 (1.5%) | 376,805,000 (1.1%) | 375,324,000 (0.7%) |
| *S. cer* | 12,157,105 | 11,872,864 (2.3%) | 12,222,224 (0.5%) | 12,180,500 (0.2%) | 12,186,800 (0.2%) |

Naive method directly uses total number of *k*-mers divided by the major peak value in the *k*-mer frequency curve. % of difference = $\frac{|estimated\ genome\ size - genome\ size|}{genome\ size}$.

**Table 6. Genuine datasets of recently published papers used in this paper**

| Sample | Species | Illumina accession | PacBio accession | Citation | Reference Genome Contig N50 Size |
|---|---|---|---|---|---|
| *C. nig* | *Caenorhabditis nigoni* | SRX3302872 | SRX3302873 | (10) | nigoni.pc_2016.07.14 3.3 Mb |
| *D. zib* | *Durio zibethinus* (fruit durian) | SRX3204603 | SRX3185921 | (11) | Duzib1.0 549,783 |
| *T. mul* | *Taenia multiceps* (tapeworm) | SRX1531874, SRX1626286, SRX1626287, SRX1626288 | SRX1531851, SRX3851532, SRX3851533, SRX3856587 | (12) | Gns01 756,412 |
| *O. tho* | *Oropetium thomaeum* | SRX1078264~ SRX1078287 | SRX1055085 | (13) | GCA_001182835.1 2,386,382 |
| *M. mic* | *Mikania micrantha* | SRR8835135~ SRR8835137 | SRR8834228~ SRR8834566 | (14) | ASM936387v1 1.35 Mb |
| *P. mil* | *Panicum miliaceum* (broomcorn millet) | SRX3628188 | SRX3628187 | (15) | Pm_0390_v2 368,640 |
| *C. kan* | *Cinnamomum kanehirae* (Stout camphor tree) | SRX4287864 | SRX4287873 | (16) | ASBRC_Ckan_1.0 498,920 |
| *P. pec* | *Pyrocoelia pectoralis* (firefly) | SRX3048410 | SRX3048411 | (17) | NA 3.04 Mb |
| *S. sin* | *Sillago sinica* (Chinese sillago) | SRX3907118, SRX3907120~ SRX3907123 | SRX3907119, SRX3907114~ SRX3907117 | (18) | NA 2.6 Mb |
| *C. ace* | *Chaenocephalus aceratus* (Antarctic blackfin icefish) | ERR1473912~ ERR1473914 | SRR6942631, SRR6942632 | (19) | NA 1.5 Mb |
| *A. aeg* | *Aedes aegypti* (yellow fever mosquito) | SRX3231347 | SRX2998003~ SRX2998178 | (20) | AaegL5.0 11,758,062 |
| *H. sap* | *Homo sapiens* (human) | | CCS dataset | https://downloads.pacbcloud.com/public/dataset/HG002_SV_and_SNV_CCS/consensusreads/ | |

NA: not available.

**Table 7. Genuine genome datasets in recently published papers**

| Species | No. of PacBio reads | No. of PacBio bases | PacBio mean read length | No. of Illumina reads | No. of Illumina bases | Illumina mean read length |
|---|---|---|---|---|---|---|
| *C. nig* | 1,480,550 | 12,538,321,717 | 8,469 | 272,260,990 | 25,685,989,344 | 94.35 |
| *D. zib* | 6,063,402 | 42,143,820,463 | 6,951 | 1,094,557,520 | 160,265,744,253 | 146.40 |
| *T. mul* | 3,451,155 | 31,470,790,545 | 8,224 | 172,170,252 | 39,496,161,611 | 228.59 |
| *O. tho* | 1,400,150 | 18,022,966,707 | 12,872 | 507,166,688 | 65,639,198,898 | 182.43 |
| *M. mic* | 25,534,574 | 228,541,311,964 | 8,950 | 852,005,706 | 122,446,081,668 | 143.72 |
| *P. mil* | 28,273,555 | 183,533,668,200 | 6,491 | 251,494,448 | 55,989,533,837 | 222.63 |
| *C. kan* | 7,916,575 | 60,745,458,308 | 7,673 | 171,824,407 | 21,450,946,075 | 124.84 |
| *P. pec* | 6,092,375 | 57,705,903,837 | 9,472 | 180,897,026 | 25,504,841,251 | 140.99 |
| *S. sin* | 3,439,159 | 27,250,347,086 | 7,924 | 568,454,060 | 79,746,400,041 | 140.29 |
| *C. ace* | 8,871,265 | 89,163,727,061 | 10,051 | 130,100,998 | 17,292,068,321 | 132.91 |
| *A. aeg* | 39,709,006 | 315,459,658,121 | 7,944 | 252,192,242 | 24,953,086,202 | 98.94 |

**Table 8. Genome size estimation for genuine datasets**

| Sample | Reported genome size | Naive method | | GCE method | |
|---|---|---|---|---|---|
| | | Corrected PacBio (% of difference) | Illumina (% of difference) | Corrected PacBio (% of difference) | Illumina (% of difference) |
| *C. nig* | 130,000,000 | 129,847,833 (0.1%) | 173,304,300 (33.3%) | 126,212,000 (2.9%) | 176,693,000 (35.9%) |
| *D. zib* | 738,000,000 | 692,717,342 (6.1%) | 861,771,670 (16.8%) | 596,672,000 (19.2%) | 946,379,000 (28.2%) |
| *T. mul* | 250,000,000 | 263,022,498 (5.2%) | 251,454,226 (0.6%) | 252,338,000 (0.9%) | 241,526,000 (3.3%) |
| *O. tho* | 245,000,000 | 229,612,886 (6.3%) | 286,456,261 (16.9%) | 232,118,000 (5.3%) | 291,244,000 (18.9%) |
| *M. mic* | 1,800,000,000 | 1,470,163,083 (18.3%) | 1,781,168,072 (1.0%) | 1,519,690,000 (15.6%) | 1,775,780,000 (1.3%) |
| *P. mil* | 923,000,000 | 819,794,391 (11.2%) | 889,633,569 (3.6%) | 811,225,000 (12.1%) | 887,807,000 (3.8%) |
| *C. kan* | 823,700,000 | 790,968,375 (4.0%) | 1,024,046,904 (24.3%) | 736,902,000 (10.5%) | 1,012,430,000 (22.9%) |
| *P. pec* | 785,000,000 | 721,623,850 (8.1%) | 737,064,735 (6.1%) | 1,093,320,000 (39.3%) | 1,290,250,000 (64.4%) |
| *S. sin* | 521,500,000 | 503,420,206 (3.5%) | 530,450,338 (1.7%) | 479,059,000 (8.1%) | 397,766,000 (23.7%) |
| *C. ace* | 1,100,000,000 | 993,316,155 (9.7%) | 1,068,609,663 (2.9%) | 930,577,000 (15.4%) | 953,663,000 (13.3%) |
| *A. aeg* | 1,250,000,000 | 1,234,884,788 (1.2%) | 1,304,219,067 (4.3%) | 1,416,450,000 (13.3%) | 1,221,250,000 (2.3%) |

Note: *A. aeg* and *P. pec* have very high heterozygous rate, thus we run GCE with "-H 1 -c peak". Here, "peak" means homogenous peak of *k*-mer frequency curve. Naive method directly uses total number of *k*-mers divided by the major peak value in the *k*-mer frequency curve. For real datasets, accurate genome size could not be known, and we just took the estimation in published papers as a comparison. % of difference = $\frac{|estimated\ genome\ size - reported\ genome\ size|}{reported\ genome\ size}$.

## 3. Supplementary figures

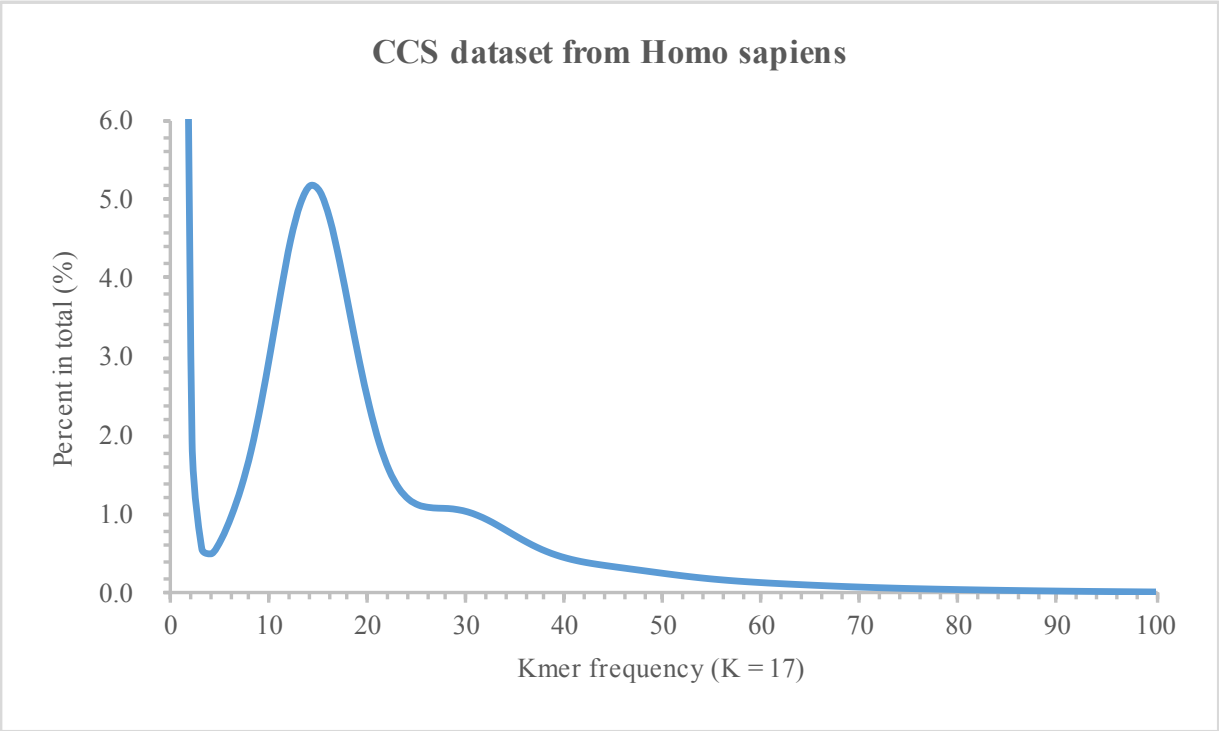

**Figure 1. The distribution of *k*-mer frequency for CCS dataset of human genome.** The peak frequency is 14 and the estimated genome size is 3,111,052,276.

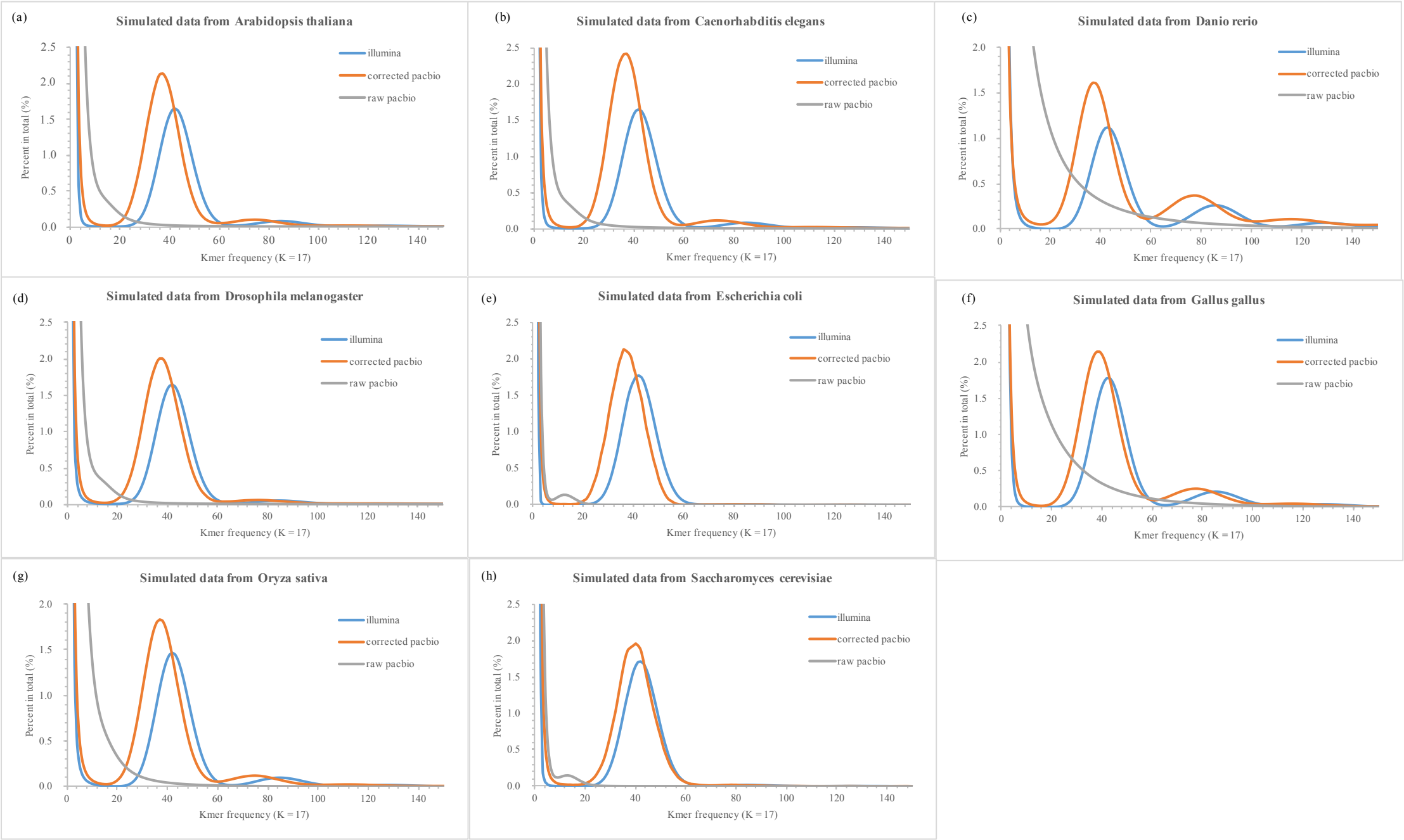

**Figure 2. The curves of *k*-mer frequency in model genomes from simulated illumina, raw pacbio and corrected pacbio data.** (a) The distributions of *k*-mer frequency of *Arabidopsis thaliana* from illumina, raw pacbio and corrected pacbio data respectively. (b) The distributions of *k*-mer frequency of *Caenorhabditis elegans* from illumina, raw pacbio and corrected pacbio data respectively. (c) The distributions of *k*-mer frequency of *Danio rerio* from illumina, raw pacbio and corrected pacbio data respectively. (d) The distributions of *k*-mer frequency of *Drosophila melanogaster* from illumina, raw pacbio and corrected pacbio data respectively. (e) The distributions of *k*-mer frequency of *Escherichia coli* from illumina, raw pacbio and corrected pacbio data respectively. (f) The distributions of *k*-mer frequency of *Gallus Gallus* from illumina, raw pacbio and corrected pacbio data respectively. (g) The distributions of *k*-mer frequency of *Oryza sativa* from illumina, raw pacbio and corrected pacbio data respectively. (h) The distributions of *k*-mer frequency of *Saccharomyces cerevisiae* from illumina, raw pacbio and corrected pacbio data respectively.

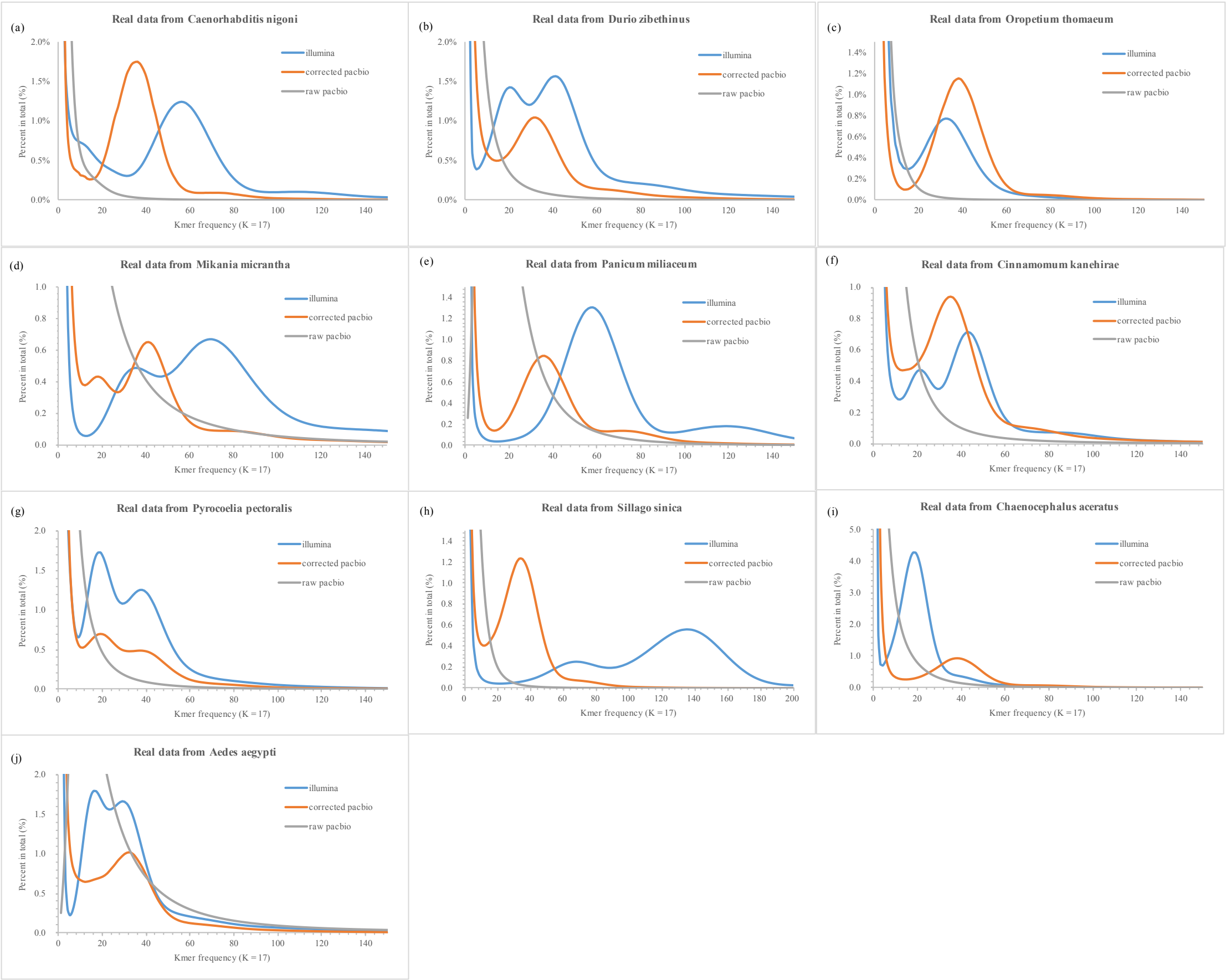

**Figure 3. The curves of *k*-mer frequency in genuine genomes from illumina, raw pacbio and corrected pacbio data.** (a) The distributions of *k*-mer frequency of *Caenorhabditis nigoni* from illumina, raw pacbio and corrected pacbio data respectively. (b) The distributions of *k*-mer frequency of *Durio zibethinus* from illumina, raw pacbio and corrected pacbio data respectively. (c) The distributions of *k*-mer frequency of *Oropetium thomaeum* from illumina, raw pacbio and corrected pacbio data respectively. (d) The distributions of *k*-mer frequency of *Mikania micrantha* from illumina, raw pacbio and corrected pacbio data respectively. (e) The distributions of *k*-mer frequency of *Panicum miliaceum* from illumina, raw pacbio and corrected pacbio data respectively. (f) The distributions of *k*-mer frequency of *Cinnamomum kanehirae* from illumina, raw pacbio and corrected pacbio data respectively. (g) The distributions of *k*-mer frequency of *Pyrocoelia pectoralis* from illumina, raw pacbio and corrected pacbio data respectively. (h) The distributions of *k*-mer frequency of *Sillago sinica* from illumina, raw pacbio and corrected pacbio data respectively. (i) The distributions of *k*-mer frequency of *Chaenocephalus aceratus* from illumina, raw pacbio and corrected pacbio data respectively. (j) The distributions of *k*-mer frequency of *Aedes aegypti* from illumina, raw pacbio and corrected pacbio data respectively.

## Supplementary References